%% file: main.tex
\documentclass[10pt, conference]{IEEEtran}
\IEEEoverridecommandlockouts
\usepackage{cite}
\usepackage{amsmath,amssymb,amsfonts, dsfont}

\usepackage{enumerate}
\usepackage{bbm}
\usepackage{algorithm2e}
\usepackage{bm}
\usepackage{algorithmic}
\usepackage{graphicx}

\makeatletter
    \let\MYcaption\@makecaption
\makeatother
    \usepackage[font=footnotesize]{subcaption}
\makeatletter
    \let\@makecaption\MYcaption
\makeatother

\usepackage{tikz}
\usetikzlibrary{shapes}
\tikzset{>=latex}

\usepackage{textcomp}
\usepackage{xcolor}
\def\BibTeX{{\rm B\kern-.05em{\sc i\kern-.025em b}\kern-.08em
    T\kern-.1667em\lower.7ex\hbox{E}\kern-.125emX}}
    \input{macros}
\begin{document}

\bstctlcite{IEEEexample:BSTcontrol}

\title{Age-Aware CSI Acquisition of a Finite-State Markovian Channel\\
\thanks{This work is a joint contribution by members of work item 205 on “6G Radio Access” of the one6G association and it has been supported in part by the European Union’s Horizon Europe MSCA-DN programme through the SCION Project under Grant Agreement No. 101072375.}
\thanks{This work has been supported in part by the Swedish Research Council (VR), ELLIIT, the Graduate School in Computer Science (CUGS), and the European Union's Horizon Europe research and innovation programme under the Marie Skłodowska-Curie Grant Agreement No 101131481 (SOVEREIGN).}
}

\author{
\IEEEauthorblockN{Onur Ayan}
\IEEEauthorblockA{\textit{Huawei Technologies} \\
Munich, Germany \\
onur.ayan@huawei.com}
\and
\IEEEauthorblockN{Jiping Luo}
\IEEEauthorblockA{\textit{Link\"{o}ping University} \\
Link\"{o}ping, Sweden \\
jiping.luo@liu.se}
\and
\IEEEauthorblockN{Xueli An}
\IEEEauthorblockA{\textit{Huawei Technologies} \\
Munich, Germany \\
xueli.an@huawei.com}
\and
\IEEEauthorblockN{Nikolaos Pappas}
\IEEEauthorblockA{\textit{Link\"{o}ping University} \\
Link\"{o}ping, Sweden \\
nikolaos.pappas@liu.se}
}

\maketitle

\begin{abstract}
The Age of Information (AoI) has emerged as a critical metric for quantifying information freshness; however, its interplay with channel estimation in partially observable wireless systems remains underexplored. This work considers a transmitter-receiver pair communicating over an unreliable channel with time-varying reliability levels. The transmitter observes the instantaneous link reliability through a channel state information acquisition procedure, during which the data transmission is interrupted. This leads to a fundamental trade-off between utilizing limited network resources for either data transmission or channel state information acquisition to combat the channel aging effect. Assuming the wireless channel is modeled as a finite-state Markovian channel, we formulate an optimization problem as a partially observable Markov decision process (POMDP), obtain the optimal policy through the relative value iteration algorithm, and demonstrate the efficiency of our solution through simulations. To the best of our knowledge, this is the first work to aim for an optimal scheduling policy for data transmissions while considering the effect of channel state information aging.
\end{abstract}


\input{introduction}
\input{system_model}

\input{policy_design}
\input{evaluation}

\section{Conclusions and Future Work}
The freshness of a real-time process is captured by the AoI metric, which has been considered an application layer metric in most of the existing literature. This work focuses on the relationship between AoI and channel state information, a topic that has not been sufficiently explored in the existing literature. In particular, we consider a multi-state Markov channel model to represent the dynamic nature of a wireless medium. We formulate a partially observable Markov decision process to determine the optimal policy for deciding when to update the channel measurement and when to utilize network resources for data transmission. Our results show that the proposed solution, obtained through the relative value iteration algorithm, outperforms the considered benchmarking policies and successfully maximizes the long-term average reward.

This work presents the results of our initial investigations in a new line of work, where the focus is set on applying the AoI metric for in-network processes. These can be channel information, as in this paper, or network-generated online data, such as in the case of in-network sensing. An immediate extension of this work is considering a more complex reward model for successful data transmissions, i.e., $R_{suc}$, such as the application layer AoI or its derivatives. Moreover, the state and action model can be detailed further, where the target reliability selection at the packet level depends on the channel age.



\bibliographystyle{IEEEtran}
\bibliography{references}

\end{document}

%% file: introduction.tex
\section{Introduction} 
\label{sec:introduction}

    Age of information (AoI) quantifies the freshness of information, and it has been widely used as a metric and tool in various application scenarios \cite{kosta2017age, yates2021ageSurvey, sun2022age, pappas2023age}. For over a decade, researchers have been investigating problems related to AoI in various settings for a wide range of applications. While the initial works focused on the theoretical characterization of AoI in queueing systems\cite{kaul2012how}, particularly in recent years, some existing works have conducted practical studies that consider AoI and its variants in networked control systems~\cite{ayan2022semantics, jiping2024MobiHoc, jiping2025TIT}. 

    Maintaining the information fresh at a monitor observing a dynamic process, typically to perform a particular task or make decisions, plays a crucial role in the accuracy and efficiency of the resulting actions. Among the well-known scenarios where AoI is a relevant tool and metric, such as in the example of environmental monitoring and networked cyber-physical control, an interesting use case in which the information freshness is important is the problem of channel state estimation in wireless systems, which has not been sufficiently studied in the existing literature. The regular acquisition of channel state information (CSI) enhances the overall system design efficiency, as CSI is utilized for user scheduling, selecting the proper modulation and coding scheme, and determining the optimal timing for data transmission.

    The acquisition of CSI, which we also refer to as channel probing throughout this work, is typically performed through the transmission of a (known) reference signal followed by subsequent signaling. Such a procedure implies that a specific portion of the available network resources (i.e., time and frequency) is allocated to perform the channel probing. While having up-to-date information about the channel state is helpful in scheduling the data transmissions, allocating a large portion of the available limited resources to channel probing is counterproductive for achieving higher throughput. This means there is an interesting trade-off between attempting transmissions when the channel conditions are sufficiently good for the packets to be successfully decoded by the receiver, while occasionally observing the channel state to identify the best moment to transmit. 

    The effects of outdated CSI on communication performance have been studied in the literature. \cite{teng2017effect} investigates how the CSI aging phenomenon affects the handover decisions in ultra-dense networks. The authors show in their paper that handover decisions are susceptible to imperfect knowledge of CSI, affecting handover failure probability. In \cite{bai2022covert}, the authors assume that only an outdated CSI can be obtained and show that the covert communication performance deteriorates monotonically as the staleness of the CSI increases. Although not applicable to wireless communications, \cite{stamatakis2023optimizing, jiping2025wearing} are relevant prior works focusing on an interesting channel model. Specifically, the authors consider a physical channel that degrades over time, i.e., its quality deteriorates with each utilization, typically observed in quantum communications. They consider AoI as the key metric to capture application layer performance and study an AoI minimization problem. In \cite{ayan2024optimal}, the authors assume a two-state Markov channel model (i.e., Gilbert-Elliott model) and study a finite horizon problem to maximize the (control) performance of feedback control applications. As an existing similarity between \cite{ayan2024optimal} and this work, one could mention that \cite{ayan2024optimal} employs the prediction of future link reliability based on the Gilbert-Elliott model. However, in \cite{ayan2024optimal}, the current channel state is always available and, thus, does not consider any aging of the CSI. Moreover, \cite{chen2021scheduling} studies an application layer cost minimization problem where the channel reliability level is not perfectly known; hence, it must be estimated. They consider a fixed error probability for incorrectly estimating the channel reliability level, which is independent of the designed policy for the application layer cost minimization problem. This differs from our system model, where the reliability estimation performance closely depends on how channel resources are utilized, e.g., for data transmission or channel state measurement.
    
    The closest works to ours are \cite{costa2015fsmc} and \cite{costa2015on}, focusing on the age of channel state information as the primary metric. \cite{costa2015on} introduces a mathematical framework to study the relationship between the channel feedback period and the channel estimation performance in the presence of feedback cost. In that work, the authors assume a two-state Markov channel model, i.e., a Gilbert-Elliott model, and determine the criteria for when to estimate the ``good'' or ``bad'' states. Similarly, \cite{costa2015fsmc} conducts a more comprehensive analysis that extends \cite{costa2015on} and investigates the relationship between the expected reward and the channel. However, neither \cite{costa2015fsmc} nor \cite{costa2015on} aims to minimize the expected reward in the system, which is the key difference between their work and ours. In fact, \cite{costa2015fsmc} mentions the design of such a policy by treating the problem as a partially observable Markov decision process (POMDP), which, to the best of our knowledge, is still missing in the existing literature. 
    
In this work, we study the problem of AoI-aware scheduling of CSI acquisition. In particular, we find an optimal policy that determines when to probe the channel, aiming to strike a balance between data transmission and channel estimation accuracy. Therefore, we design a reward model that penalizes the staleness of channel state information and rewards successful data transmission. Our results show that the proposed solution, obtained through the relative value iteration algorithm, outperforms the considered benchmarking policies and maximizes the long-term average reward.

The paper is structured as follows. Section~\ref{sec:system_model} presents the system model and defines the age of channel state information. Section~\ref{sec:policy_design} introduces the reward model, formulates the optimization problem, and proposes an AoI-aware optimal policy. Finally, Section~\ref{sec:evaluation} presents the numerical results.

%% file: system_model.tex
\section{System Model}


\label{sec:system_model}
\subsection{System Description}
\begin{figure}
    \centering
    \includegraphics[width=\linewidth]{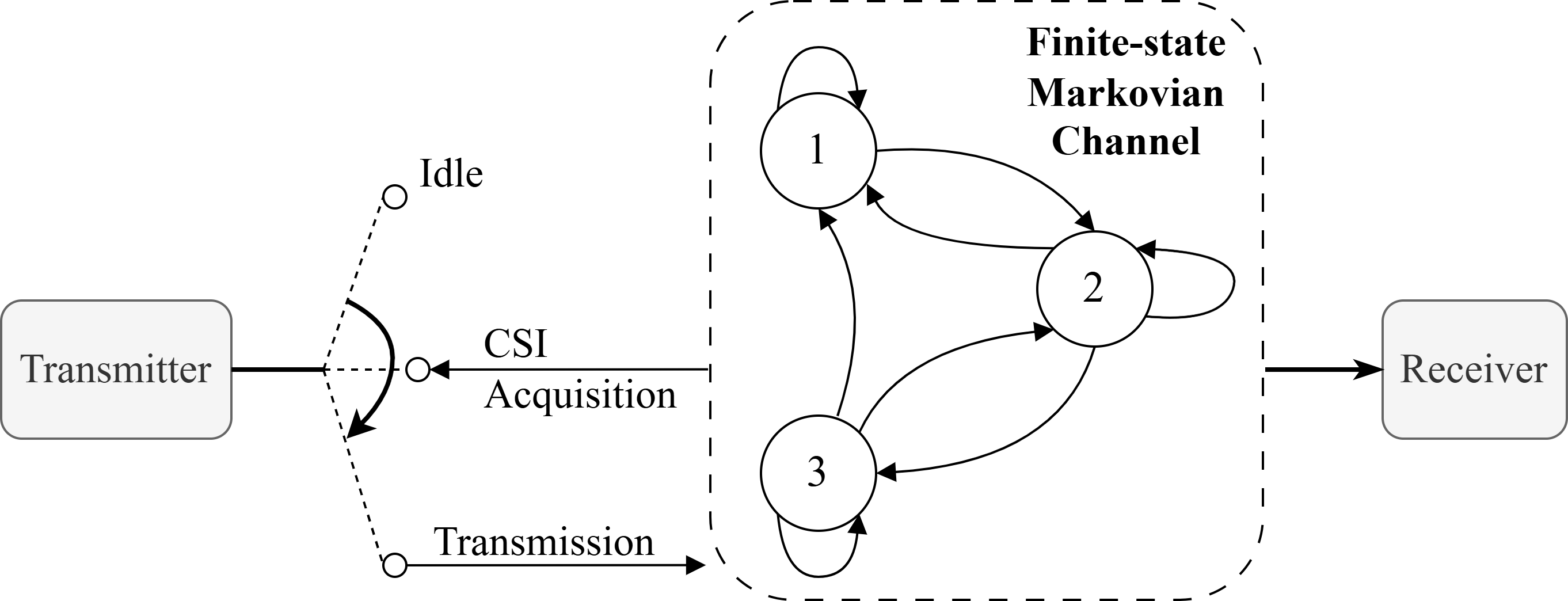}
    \caption{Considered scenario with the transmitter-receiver pair connected via a finite-state Markovian channel. The policy at the transmitter decides between three actions, namely, idle, CSI acquisition, and data transmission.}
    \label{fig:scenario}
\end{figure}

We consider a transmitter-receiver pair communicating over an unreliable channel to monitor a source remotely, as depicted in Fig. \ref{fig:scenario}. We assume that the source generates a new packet carrying the latest status update at every time slot $t$, which is the finest granularity of time in our system model. We assume that a packet transmission at a time step $t$ occupies the entire slot and ends before the next slot, $t+1$, begins.

The behavior of the wireless link is modeled as an $M$-state Markov chain with the transition matrix $\bm{P} \in \mathbb{R}^{M \times M}$. Each channel state $s[t] \in \{ 0, 1, \dots, M-1\}$ is associated with a certain time-invariant reliability level $r(s)$, with the mapping function $r: s \mapsto [0, 1]$. Similar to the source process, the channel state is updated every time slot. Finite-state Markovian channels have been widely used in the state-of-the-art to model packet loss in real-time networks, e.g., in \cite{hasslinger2008gilbert, bildea2015link} for Gilbert-Elliott model (i.e., $M=2$). While the
two-state model has been shown to be fairly accurate to
represent Rayleigh-fading channels when the channel quality
does not vary dramatically over time, a model with more than two states is recommended for dynamic scenarios \cite{wang1995finite}.

The transmitter only observes the CSI, denoted by $s[t]$ in time slot $t$, through an acquisition procedure that allows the transmitter to measure it perfectly. We assume that the CSI acquisition procedure occupies an entire slot, as in the case of data transmission. Moreover, the CSI acquisition and data transmission cannot be performed simultaneously. The time-invariant mapping between a channel state and the link reliability is available to the transmitter in that state, i.e., $r(s)$.

The transmitter employs a policy $\pi_a$ that performs one of three actions at every decision epoch (i.e., at the beginning of each time slot $t$): i) remain ``idle'' by neither transmitting nor acquiring the CSI; ii) initiate the CSI acquisition procedure that allows it to observe $s[t]$ perfectly, thus, measure the link reliability $r(s[t])$;  iii) attempt a new packet transmission, which will be successfully received by the monitor with a probability of $r(s[t])$ depending on the current channel state $s[t]$. The action taken by the transmitter in slot $t$ is denoted by $a[t] \in \mathcal{A}$, with $\mathcal{A} = \{I, C, T\}$. Here, the feasible actions $I$, $C$, and $T$ correspond to the options i), ii), and iii), respectively\footnote{The reader can also assume that $I$ stands for idle, $C$ stands for CSI acquisition, and $T$ stands for transmission.}. Moreover, we consider an energy cost model for each action. The transmission action incurs a particular energy cost $\epsilon_T$, whereas the CSI acquisition costs $\epsilon_C$ with $0< \epsilon_C < \epsilon_T$. The idle action is cost-free, i.e., $\epsilon_I = 0$.

\begin{figure}[t]
    \centering
    \scalebox{0.90}{\input{images/AoCSI.tex}}
    \caption{The evolution of the age of CSI. The transmitter probes the channel at decision epochs $t_1$, $t_4$, and $t_5$, i.e., $a[t_1] = a[t_4] = a[t_5] = C$. The age resets to $1$ at epochs $t_2, t_5$, and $t_6$, as characterized by \eqref{eq:channel_age}.}
    \label{fig:AoCSI}
\end{figure}
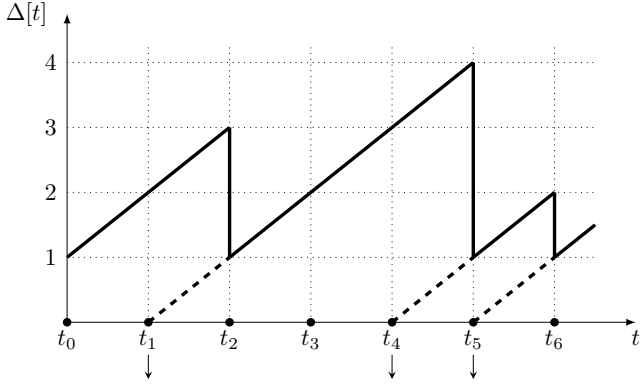

We introduce a new variable to quantify the freshness of the CSI at the transmitter. The \textit{Age of Channel State Information (AoCSI)} is defined as the number of time slots that have elapsed since the transmitter's most recent CSI acquisition until time slot $t$. The AoCSI $\Delta[t] \in \mathbb{Z}^{+}$ is defined as:
\begin{equation}
    \Delta[t] \triangleq t - \max_{\tau \in \mathbb{N}} \{ \tau <  t : a[\tau] = C\}. 
\end{equation}
This implies that the most recent observation of the channel state available to the transmitter is $s[t - \Delta[t]]$. Fig.~\ref{fig:AoCSI} shows the evolution of the AoCSI. We note that the smallest AoCSI value can take is $1$ at any decision epoch $t$. For instance, at decision epoch $t_1$, the transmitter decides to probe the channel based on outdated information with an age of $2$. Since the CSI acquisition procedure occupies an entire slot, the AoCSI will reset to $1$ at decision epoch $t_2$. Formally, we have
\begin{align}
    \Delta[t+1] = \begin{cases}
        1, &\textrm{if}~a[t] = C,\\
        \Delta[t] + 1, &\textrm{otherwise}.
    \end{cases}
    \label{eq:channel_age}
\end{align}
Thus, we have a \emph{partially observable system}, as the transmitter does not have exact knowledge of the underlying process. 
\begin{figure*}[t]
    \begin{subfigure}{0.49\linewidth}
        \centering
        \includegraphics[width=0.95\linewidth]{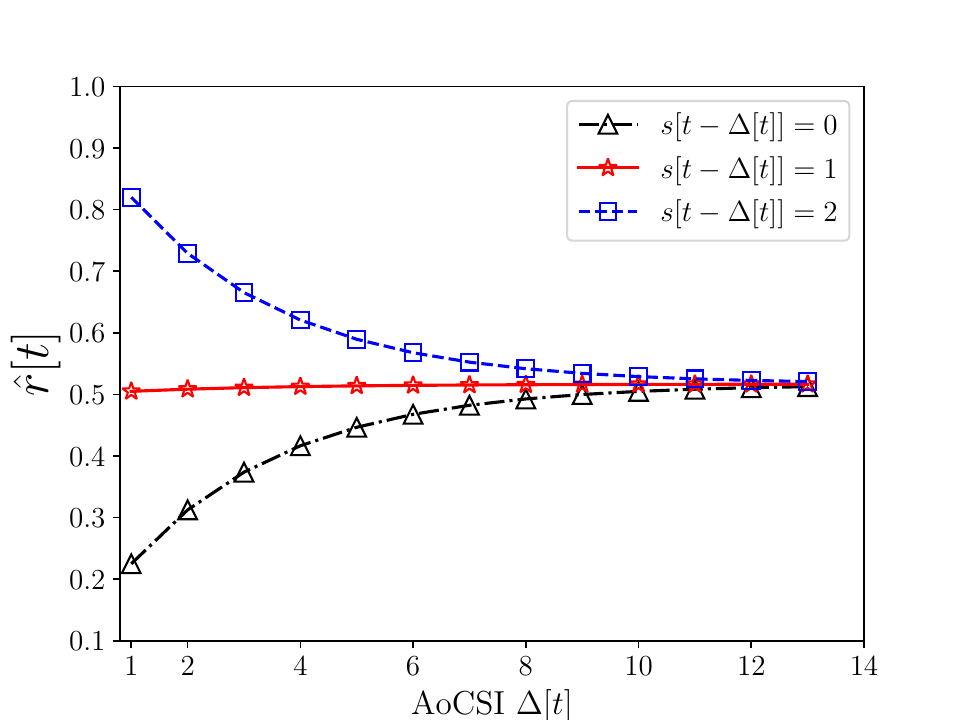}
        \caption{Expected link reliability.}
    \label{fig:expected-reliability}
    \end{subfigure}
    \begin{subfigure}{0.49\linewidth}
        \centering
        \includegraphics[width=0.95\linewidth]{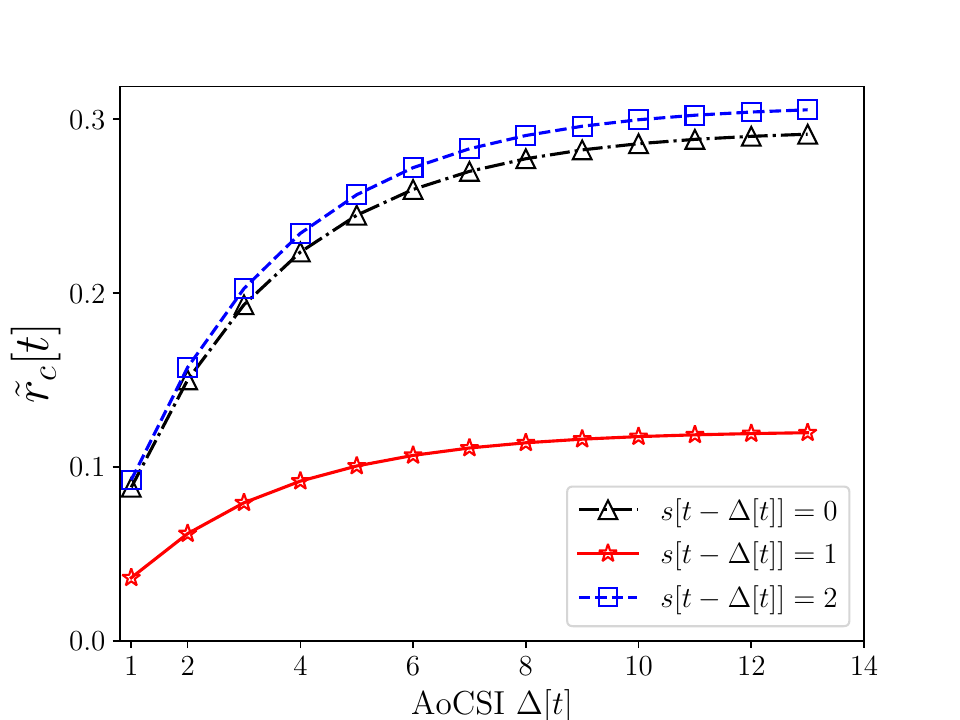}
        \caption{MSE of link reliability.}
    \label{fig:expected-mse-reliability}
    \end{subfigure}  
    \vspace{-0.1in}
    \caption{Age-dependent dynamics of expected link reliability and expected reliability estimation error in squared form illustrated in Fig. \ref{fig:expected-reliability} and \ref{fig:expected-mse-reliability}, respectively. Here, the $3 \times 3$ transition matrix is symmetric with diagonal elements being $P_{i,i} = 0.8$ and off-diagonal elements being $P_{i, j} = 0.1$. The link reliabilities for each channel state are $r(s) \in \{0.1, 0.5, 0.95\}$.}
    \label{fig:evolution of the rewards}
\end{figure*}

Let $\bm{z}[t] \triangleq \left[ \Delta[t] \quad s[t - \Delta[t]]\right]^T$ represent the overall system state available to the transmitter at decision epoch $t$. We can formulate the time behavior of $\bm{z}[t]$ depending on the taken action as follows. For the case when $a[t] \in \{I, T\}$, i.e., the transmitter decides for a transmission or an idle slot, the system state at decision epoch $t+1$ can be determined as:
\begin{align}
    \bm{z}[t+1] = \left[ \Delta[t] + 1 \quad s[t - \Delta[t] - 1] \right]^T.
\end{align}
If a probing action is taken, i.e., $a[t] = C$, the transmitter acquires the most recent channel information, aged by only $1$ slot, i.e.,
\begin{equation}
    \bm{z}[t+1] = \left[ 1 \quad s[t - 1]\right]^T.
\end{equation}

\subsection{Channel state and reliability estimation}
As we are dealing with a partially observable system, the transmitter must rely on the latest observation in order to estimate the current state\footnote{An alternative approach would be to assume that the channel state remains constant between two consecutive CSI acquisitions, which is also called zero-order hold in the literature.}. Nevertheless, since our channel model is characterized by an $M$-state Markov chain with the transition matrix $\bm{P}$, this allows us to obtain the probability of being at a particular channel state, given the latest observation:
\begin{equation}
    p_{\sigma_1, \sigma_0}^{(\delta)} \triangleq \Pr[s[t] = \sigma_1 \mid \bm{z}[t] = \left[ \delta \quad \sigma_0 \right]^T].
    \
    \label{eq:state_probability}
\end{equation}

It is known that \eqref{eq:state_probability} can be calculated using the Chapman-Kolmogorov equations \cite[Ch. 4.2]{ross2014introduction}, which simply state that the $n$-step transition matrix of a Markov chain can be obtained by taking the $n$-th matrix power, i.e., $\bm{P}^n$. More specifically, the probability $p_{\sigma_1, \sigma_0}^{(\delta)}$ from \eqref{eq:state_probability} is the entry located in the $\sigma_1$-th row and $\sigma_0$-th column of the $\Delta[t]$-step transition matrix $\bm{P}^{\Delta[t]}$. Since the relationship between a channel state and the link reliability is known, it is straightforward to calculate the expected link reliability given the latest CSI acquisition:
\begin{align}
    \hat{r}[t] \triangleq \mathbb{E}\left[ r(s[t]) \mid \bm{z}[t] = \left[ \delta \quad \sigma_0 \right]^T \right] 
    =  \sum_{\sigma_1}  p_{\sigma_1, \sigma_0}^{(\delta)} r(\sigma_1).
    \label{eq:exp_reliability}
\end{align}
Fig.~\ref{fig:expected-reliability} shows the evolution of the expected link reliability as a function of AoCSI for a three-state Markov channel model, i.e., $M = 3$. Each of the three trajectories show the expected reliability at $\Delta[t]$ given that the last observed channel state, i.e,. $s[t - \Delta[t]]$. The figure shows the convergence of the three curves around the steady-state value for large AoCSI, which lies at $\approx 0.517$ for the selected parameters. The convergence phenomenon of the expected reliability shows that the most recent information that was acquired about the channel state $\Delta[t]$ slots ago becomes obsolete  and does not provide any useful information about the current channel state.


%% file: images/AoCSI.tex
\begin{tikzpicture}[scale=1.2]
\draw[->] (0,0) -- (7,0) node[anchor=north] {$t$};
\draw[->] (0,0) -- (0,3.8);
\draw (-0.5, 3.8) node {$\Delta[t]$};
\draw (-0.2, 0.8) node {$1$};
\draw (-0.2, 1.6) node {$2$};
\draw (-0.2, 2.4) node {$3$};
\draw (-0.2, 3.2) node {$4$};

\fill (0,0)  circle[radius=1.5pt];
\fill (1,0)  circle[radius=1.5pt];
\fill (2,0)  circle[radius=1.5pt];
\fill (3,0)  circle[radius=1.5pt];
\fill (4,0)  circle[radius=1.5pt];
\fill (5,0)  circle[radius=1.5pt];
\fill (6,0)  circle[radius=1.5pt];

\draw	(0,0) node[anchor=north] {$t_0$}
        (1,0) node[anchor=north] {$t_1$}
		(2,0) node[anchor=north] {$t_2$}
		(3,0) node[anchor=north] {$t_3$}
		(4,0) node[anchor=north] {$t_4$}
		(5,0) node[anchor=north] {$t_5$}
        (6,0) node[anchor=north] {$t_{6}$};

\draw[dotted] (1,0) -- (1,3.4);
\draw[dotted] (2,0) -- (2,3.4);
\draw[dotted] (3,0) -- (3,3.4);
\draw[dotted] (4,0) -- (4,3.4);
\draw[dotted] (5,0) -- (5,3.4);
\draw[dotted] (6,0) -- (6,3.4);
\draw[dotted] (0,0.8) -- (6.5,0.8);
\draw[dotted] (0,1.6) -- (6.5,1.6);
\draw[dotted] (0,2.4) -- (6.5,2.4);
\draw[dotted] (0,3.2) -- (6.5,3.2);

\draw[line width=0.5mm]  (0,0.8) to[bend right=0] (2,2.4); 
\draw[line width=0.5mm]  (2,2.4) to[bend right=0] (2,0.8); 
\draw[dashed, line width=0.5mm]  (1,0.) -- (2,0.8); 

\draw[line width=0.5mm]  (2,0.8) to[bend left=0] (5,3.2); 
\draw[line width=0.5mm]  (5,3.2) to[bend left=0] (5,0.8); 
\draw[dashed, line width=0.5mm]  (4,0.) -- (5,0.8); 

\draw[line width=0.5mm]  (5,0.8) to[bend left=0] (6,1.6); 
\draw[line width=0.5mm]  (6,1.6) to[bend left=0] (6,0.8);
\draw[dashed, line width=0.5mm]  (5,0.) -- (6,0.8);

\draw[line width=0.5mm]  (6,0.8) to[bend left=0] (6.5,1.2); 

\draw[->,>=stealth]    (1,-0.4) -- (1,-0.7);
\draw[->,>=stealth]    (4,-0.4) -- (4,-0.7);
\draw[->,>=stealth]    (5,-0.4) -- (5,-0.7);
         
\end{tikzpicture}

%% file: policy_design.tex
\section{Optimal Policy Design Based on Age of Channel State Information}
\label{sec:policy_design}
\subsection{Reward Model}
Although the ability to successfully track the CSI plays a pivotal role in our considered scenario, our primary goal is to use the information acquired about the channel for reliable data transmission. This renders finding the right balance between the CSI acquisition and data transmission crucial and calls for a reward model that incorporates both aspects in the decision-making process.


We define a \textit{reward function} $R: \mathcal{Z} \times \mathcal{A} \mapsto \mathbb{R}$ that considers the uncertainty in channel estimation, while also rewarding a successful data transmission:
\begin{equation}
    R(\bm{z}[t], a[t]) \triangleq R_T(\bm{z}[t], a[t]) + \beta R_C(\bm{z}[t], a[t]).
\end{equation}
Here, $\beta \geq 0$ is a constant scaling factor, which balances the importance of transmission gain relative to the channel uncertainty. Specifically, a large $\beta$ may result in frequent channel probing, while a small $\beta$ places greater emphasis on data transmission. The first component $R_T$ is a function mapping to a non-negative reward only if the action is to transmit, i.e.,
\begin{equation}
    R_T(\bm{z}[t], a[t]=T) \triangleq \hat{r}[t] R_{suc} + (1 - \hat{r}[t]) R_{fail} - \epsilon_T
\end{equation}
and $R_T(\bm{z}[t], a[t]\neq T) = 0$. 
The second component, i.e., $R_C$, is a non-positive reward function penalizing the error in link reliability estimation. We select the negative mean squared error (MSE) as follows
\begin{equation}
    R_{C}[t] \triangleq  - \mathds{1}\left(a[t] \neq C\right) \Tilde{r}_c[t] - \mathds{1}\left(a[t] = C\right) \epsilon_C,
    \label{eq:channel_cost}
\end{equation}
with:
\begin{align}
    \Tilde{r}_c[t] &\triangleq \mathbb{E}\left[ \left( r[t] - \hat{r}[t] \right)^2 \mid \bm{z}[t] = \left[ \delta \quad \sigma_0 \right]^T \right] \nonumber \\
    &= \sum_{s} \Pr\left[s[t] = s \mid \bm{z}[t]\right] \left( r(s) - \hat{r}[t] \right)^2 \nonumber \\
    &= \sum_{s} p_{s, \sigma_0}^{\delta} \left( r(s) - \hat{r}[t] \right)^2.
\end{align}
$R_C$ is a non-linear function of channel age, i.e., $\Delta[t]$, while being conditioned on the latest channel observation. In addition, we note that the evolution of $\Tilde{r}_c$, hence, of $R_C$ is identical both for $a[t] = I$ and $a[t] = T$.  In particular, one can consider $R_C$ as a set of $M$ distinct time-invariant mappings from the channel age to the squared form of the expected reliability estimation error given the latest observed channel state. The non-linear relationship between the AoCSI and the MSE is shown in Fig. \ref{fig:expected-mse-reliability}. Similar to Fig. \ref{fig:expected-reliability}, one can observe the convergence of MSE for large values of $\Delta[t]$ also in Fig. \ref{fig:expected-mse-reliability}.

\subsection{Problem Statement and Optimal Policy}
Our objective is to find an admissible policy $\pi_a$ that is able to solve an infinite horizon problem by maximizing the expected reward when employed. In particular, we are interested in finding the policy that maximizes the expected reward for policy $\pi_a$ as follows:
\begin{equation}
    \max_{\pi_a} \; \; \mathbb{E}_{\pi_a} \left[\limsup_{T\rightarrow\infty}\frac{1}{T} \sum_{t=0}^{T - 1} R(\bm{z}[t], a[t]) \right].
    \label{eq:problem_statement}
\end{equation}
Problem~\eqref{eq:problem_statement} is a \emph{partially observable Markov decision process} (POMDP). In this paper, we utilize the \textit{relative value iteration (RVI)} algorithm to solve this problem, as shown in Algorithm \ref{alg:rvi}\cite{puterman1994markov}. The RVI algorithm distinguishes a selected state, i.e., the reference state denoted by $\bm{z}_{ref}$, by subtracting its value, $v(\bm{z}_{ref})$, from the value of each state, $v(\bm{z})$, after each iteration step, with $\bm{z}, \bm{z}_{ref} \in \mathcal{Z}$. The iteration continues until the difference in relative values become smaller than a threshold $\theta$, returning the $\theta$-optimal policy $\pi_{a}$ as its output.

It is evident from Algorithm \ref{alg:rvi} that the RVI iterates through the entire state space $\mathcal{Z}$ in order to update the state value. However, the $\mathcal{Z}$ is a countable infinite set due to the fact that the age of CSI, i.e., $\Delta[t]$, can take infinitely large values. Fortunately, as shown in Fig.~\ref{fig:evolution of the rewards}, the value of $\Tilde{r}$, and hence $R_C$ and $R_T$, converge as the AoCSI increases. This is because each row of $\bm{P}^{\Delta[t]}$, i.e., $\bm{P}^{\Delta[t]}_\sigma, \forall \sigma \in\{0, 1, \ldots, M-1\}$, converges to the steady-state distribution of the Markov chain. So there is no loss of optimality in truncating the AoCSI with a sufficiently large bound. As a solution, we consider an upper-bound $\Delta_{max}$ that allows us to approximate Problem~\eqref{eq:problem_statement}. 

It is important to mention that in a practical deployment scenario, the RVI algorithm will run offline (i.e., prior to deployment)~\cite{luo2025TCOM}. This means that the determined policy $\pi_a$, which provides the transmitter with the optimal action in each network state $\bm{z}[t]$, is only used as a lookup table. This property improves the practicality of our solution. The infinite-horizon problem is characterized by the $M$-state Markov channel model. In case the parameters of the channel model should change during operation, the RVI algorithm should be re-run using the updated model to obtain the optimal policy $\pi_a$.

\input{rvi_alg}

%% file: rvi_alg.tex
\SetKwComment{Comment}{/* }{ */}
\SetKwRepeat{Do}{do}{while}
\SetKwInOut{Input}{Input}\SetKwInOut{Output}{Output}
\RestyleAlgo{ruled}
\begin{algorithm}
\caption{Relative Value Iteration Algorithm}
\label{alg:rvi}
\Input{$\bm{z}_{ref} \in \mathcal{Z}$, $\theta >0$, $w_0(\bm{z})$}
\Output{$\pi_a$}
$w_0(\bm{z}_{ref}) \gets 0$; \\
$n \gets 0$; \\

\Do{$\underset{s \in \mathcal{S}}{\max}\left\{\Vert w_{n+1}(s) - w_n(s)\Vert\right\} \geq \theta$}
{
\Comment{Iterate}
\For{$\bm{z} \in \mathcal{Z}$}{
    $v_{n}(\bm{z}) \gets \underset{a \in \mathcal{A}}{\max} \left\{ R(\bm{z}, a) + \underset{\bm{z'} \in \mathcal{Z}}{\sum} \Pr[\bm{z'} \mid \bm{z}, a] \, w_n(\bm{z'})\right\};$ \\
}
\Comment{Normalize}
\For{$\bm{z} \in \mathcal{Z}$}{
 $w_n(\bm{z}) \gets v_n(\bm{z}) - v_n(\bm{z}_{ref});$ \\
}
    $n \gets n +1;$\\
}
\Comment{Obtain $\theta$-optimal policy}
$\pi_a(\bm{z}) \gets \underset{a \in \mathcal{A}}{\arg \max} \left\{ R(\bm{z}, a) + \underset{\bm{z'} \in \mathcal{Z}}{\sum} \Pr[\bm{z'} \mid \bm{z}, a] \, w_n(\bm{z'})\right\};$
\end{algorithm}

%% file: evaluation.tex
\section{Numerical Evaluation and Key Results}
\label{sec:evaluation}
\begin{figure*}[t]
    \begin{subfigure}{0.49\linewidth}
        \centering
        \includegraphics[width=0.9\linewidth]{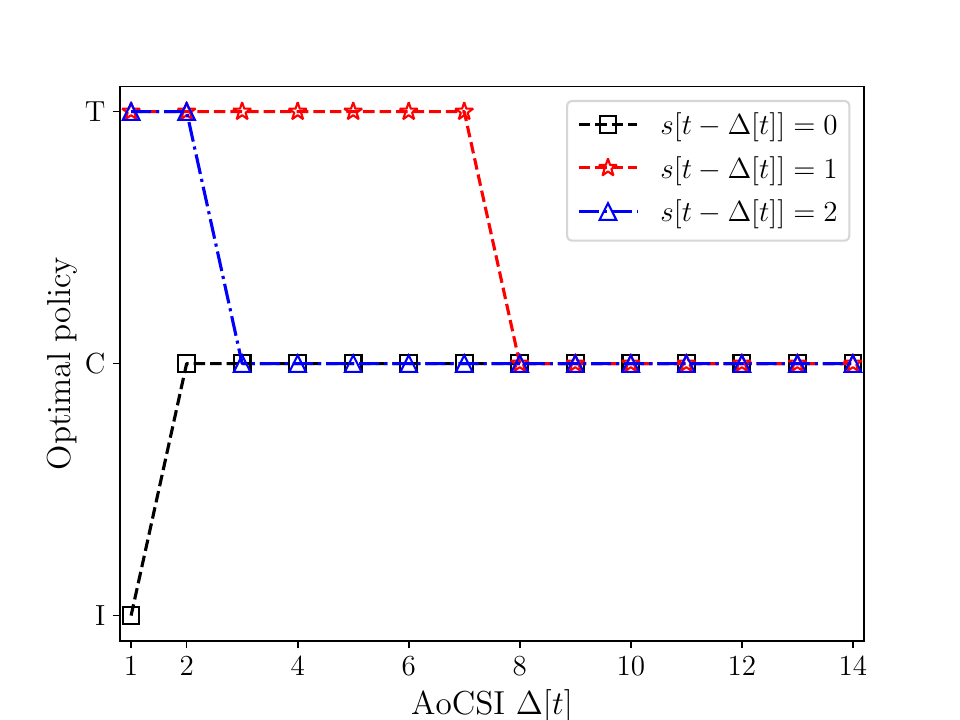}
        \caption{Optimal policy}
        \label{fig:optimal-policy}
    \end{subfigure}
    \begin{subfigure}{0.49\linewidth}
        \centering
        \includegraphics[width=0.9\linewidth]{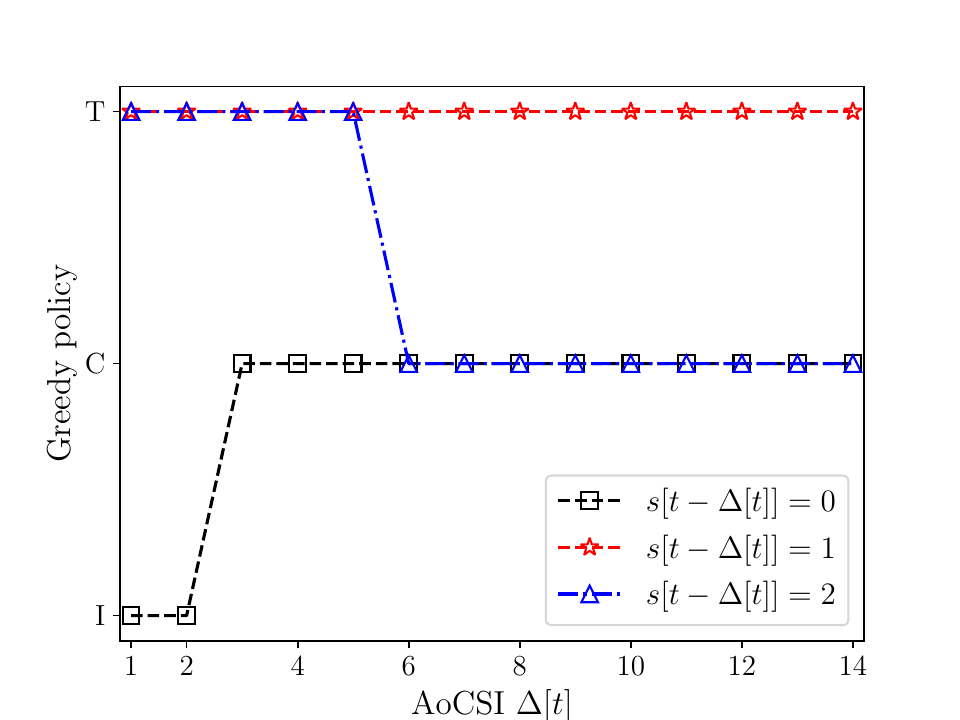}
        \caption{Greedy policy}
        \label{fig:greedy-policy}
    \end{subfigure}
    \caption{The visualization of optimal and greedy policies for a three-state Markov channel model for $\Delta_{max} = 14$. The optimal policy results in an earlier channel probing action when compared to the greedy policy. The parameters of the channel model are the same as in Fig. \ref{fig:evolution of the rewards}.}
    \label{fig:main-caption-of-performance-comparison}
\end{figure*}

The optimal policy obtained through the RVI algorithm is designed to minimize the expected average infinite horizon cost given in \eqref{eq:problem_statement}. In order to validate the optimality of our proposed solution, we conduct numerical evaluation based on simulations. For our evaluations, we consider a channel model with $M = 3$ states, where states $0, 1$ and $2$ represent the \textsc{Bad}, \textsc{Mediocre}, and \textsc{Good} channel conditions, respectively. The transition matrix $\bm{P}$ is symmetric, with $\bm{P}_{i,j} = 0.1, \forall i\neq j$, and $0.8$ otherwise. The channel reliability levels are set to $r(0) = 0.1, r(1) = 0.5, r(2) = 0.95$, which correspond to the packet success probability at each state. We select the reward for successful and failed transmissions as $R_{suc} = 1$ and $R_{fail} = 0$ and the scaling factor for MSE is selected as $\beta = 1.8$. 

We consider two benchmark solutions to compare our evaluation results with: i) the greedy policy; ii) the randomized policy. As the name suggests, the randomized policy chooses a random action out of the set of feasible actions, $\mathcal{A}$, with equal probabilities. Evidently, the randomized policy is a model-unaware, hence, an age-unaware policy. On the contrary, the greedy policy is a model-aware policy that always decides for the action maximizing the immediate reward at each state, i.e.,
\begin{equation}
    \pi_a(\bm{z}[t]) = \underset{a \in \mathcal{A}}{\arg \max} \; \; R(\bm{z}[t], a[t]).
    \label{eq:greedy_policy}
\end{equation}
The reason for considering a heuristic solution as a benchmark, such as the one characterized by \eqref{eq:greedy_policy}, is to assess the benefit of solving the infinite horizon problem optimally using the RVI algorithm, although it is comparably more complex to implement.

Fig.~\ref{fig:optimal-policy} shows the behavior of the optimal policy when the cost of channel acquisition and data transmission are $\epsilon_C = 0.3$ and $\epsilon_T = 0.4$. The first important observation to mention is that the optimal timing for the first CSI acquisition depends on the latest observed channel state. For instance, when the channel is more likely to reside in the \textsc{Bad} state, i.e., $s[t-\Delta[t]] = 0$ and $\Delta[t]$ is small, the optimal action is to remain idle and probe the channel afterwards to detect a possible change in the channel state. Conversely, when the channel is expected to be in a relatively good state, it is optimal to transmit continuously for several consecutive time slots and only probe the channel when the AoCSI exceeds a certain threshold. It is worth mentioning that the system is ``restarted" when a channel probing action occurs. What we mean is the following. When $a[t] = C$, the AoCSI is reset to zero which makes the policy follow one of the three curves depending on the observed channel state. Since we consider that a CSI acquisition resets $\Delta[t]$ to zero deterministically, the obtained policy will never exceed the threshold along the respective curve, rendering the states beyond $\Delta[t] = \delta_{th}$ unreachable in our scenario. Note that $\delta_{th} \in \{2, 3, 8\}$ in Fig.~\ref{fig:optimal-policy}.





\begin{figure}[ht]
    \centering
    \includegraphics[width=0.95\linewidth]{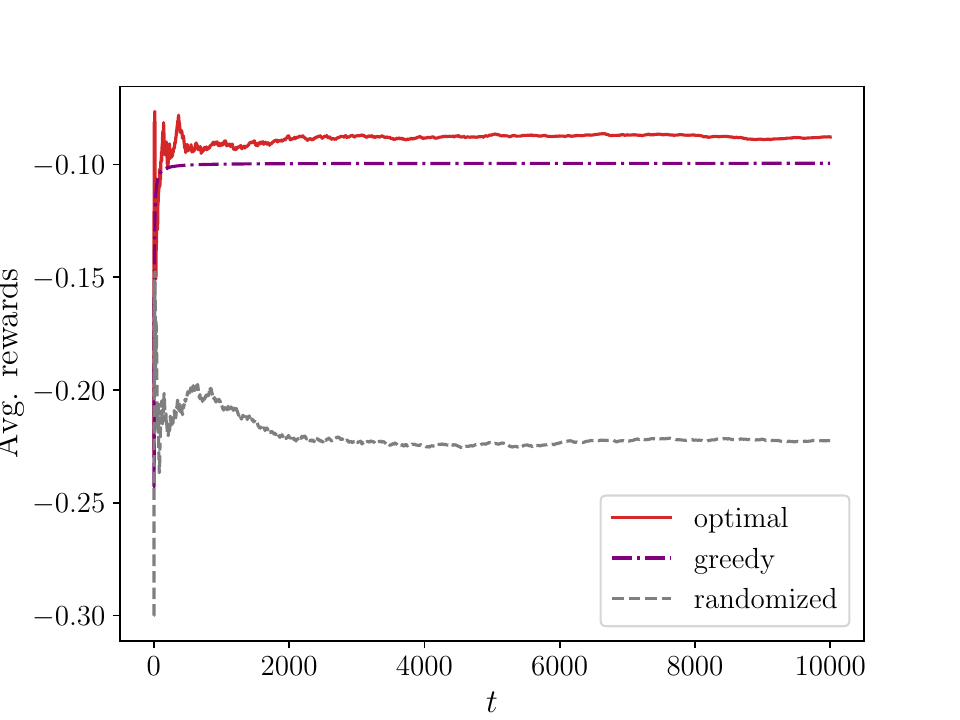}
    \caption{The long-term average reward achieved by the considered policies.}
    \label{fig:performance-comparsion}
\end{figure}

On the other hand, as shown in Fig.~\ref{fig:greedy-policy}, the greedy policy probes the channel less often. This is because the greedy policy selects the action that maximizes only the reward in the next slot. However, besides MSE reduction, channel acquisition provides fresher CSI and hence, more informed decisions by solving the POMDP. This accounts for the long-term performance gain achieved by the optimal policy in Fig.~\ref{fig:performance-comparsion}. 

Fig. \ref{fig:performance-comparsion} compares the long-term average reward achieved by different policies. As expected, the randomized policy performs significantly worse among the three considered options due to its negligence to the system knowledge. From the figure, we can observe that the greedy policy based on our reward model achieves a much higher average reward when compared to the randomized policy. However, the figure shows that the performance can be improved further when the optimal policy is obtained through the RVI algorithm introduced in Sec. \ref{sec:policy_design}.

%

%% file: main.bbl
\begin{thebibliography}{10}
\providecommand{\url}[1]{#1}
\csname url@samestyle\endcsname
\providecommand{\newblock}{\relax}
\providecommand{\bibinfo}[2]{#2}
\providecommand{\BIBentrySTDinterwordspacing}{\spaceskip=0pt\relax}
\providecommand{\BIBentryALTinterwordstretchfactor}{4}
\providecommand{\BIBentryALTinterwordspacing}{\spaceskip=\fontdimen2\font plus
\BIBentryALTinterwordstretchfactor\fontdimen3\font minus \fontdimen4\font\relax}
\providecommand{\BIBforeignlanguage}[2]{{%
\expandafter\ifx\csname l@#1\endcsname\relax
\typeout{** WARNING: IEEEtran.bst: No hyphenation pattern has been}%
\typeout{** loaded for the language `#1'. Using the pattern for}%
\typeout{** the default language instead.}%
\else
\language=\csname l@#1\endcsname
\fi
#2}}
\providecommand{\BIBdecl}{\relax}
\BIBdecl

\bibitem{kosta2017age}
A.~Kosta, N.~Pappas, and V.~Angelakis, ``Age of information: A new concept, metric, and tool,'' \emph{Foundations and Trends® in Networking}, vol.~12, no.~3, pp. 162--259, 2017.

\bibitem{yates2021ageSurvey}
R.~D. Yates, Y.~Sun, D.~R. Brown, S.~K. Kaul, E.~Modiano, and S.~Ulukus, ``Age of information: An introduction and survey,'' \emph{IEEE Journal on Selected Areas in Communications}, vol.~39, no.~5, 2021.

\bibitem{sun2022age}
Y.~Sun, I.~Kadota, R.~Talak, and E.~Modiano, \emph{Age of information: A new metric for information freshness}.\hskip 1em plus 0.5em minus 0.4em\relax Springer Nature, 2022.

\bibitem{pappas2023age}
{N. Pappas et al.}, \emph{Age of Information: Foundations and Applications}.\hskip 1em plus 0.5em minus 0.4em\relax Cambr. Univ. Press, 2023.

\bibitem{kaul2012how}
S.~Kaul, R.~Yates, and M.~Gruteser, ``Real-time status: How often should one update?'' in \emph{IEEE International Conference on Computer Communications (INFOCOM}, 2012, pp. 2731--2735.

\bibitem{ayan2022semantics}
O.~Ayan, P.~Kutsevol, H.~Y. Özkan, and W.~Kellerer, ``Semantics- and task-oriented scheduling for networked control systems in practice,'' \emph{IEEE Access}, vol.~10, pp. 115\,673--115\,690, 2022.

\bibitem{jiping2024MobiHoc}
J.~Luo and N.~Pappas, ``Minimizing the age of missed and false alarms in remote estimation of {Markov} sources,'' in \emph{Proc. 25th Int. Symp. Theory Algorithmic Found. Protoc. Des. Mobile Netw. Mobile Comput. (MobiHoc)}, Athens, Greece, Oct. 2024, pp. 381--386.

\bibitem{jiping2025TIT}
J.~Luo and N.~Pappas, ``On the cost of consecutive estimation error: Significance-aware non-linear aging,'' \emph{IEEE Transactions on Information Theory}, Jul. 2025, early access.

\bibitem{teng2017effect}
Y.~Teng, M.~Liu, and M.~Song, ``Effect of outdated csi on handover decisions in dense networks,'' \emph{IEEE Communications Letters}, vol.~21, no.~10, pp. 2238--2241, 2017.

\bibitem{bai2022covert}
J.~Bai, J.~He, Y.~Chen, Y.~Shen, and X.~Jiang, ``On covert communication performance with outdated csi in wireless greedy relay systems,'' \emph{IEEE Transactions on Information Forensics and Security}, vol.~17, pp. 2920--2935, 2022.

\bibitem{stamatakis2023optimizing}
G.~J. Stamatakis, O.~Simeone, and N.~Pappas, ``Optimizing information freshness over a channel that wears out,'' in \emph{Asilomar Conference on Signals, Systems, and Computers}, 2023, pp. 85--89.

\bibitem{jiping2025wearing}
\BIBentryALTinterwordspacing
J.~Luo, G.~Stamatakis, O.~Simeone, and N.~Pappas, ``Remote state estimation over a wearing channel: Information freshness vs. channel aging,'' 2025. [Online]. Available: \url{https://arxiv.org/abs/2501.17473}
\BIBentrySTDinterwordspacing

\bibitem{ayan2024optimal}
O.~Ayan, S.~Hirche, A.~Ephremides, and W.~Kellerer, ``Optimal finite horizon scheduling of wireless networked control systems,'' \emph{IEEE/ACM Transactions on Networking}, vol.~32, no.~2, pp. 927--942, 2024.

\bibitem{chen2021scheduling}
Y.~Chen and A.~Ephremides, ``Scheduling to minimize age of incorrect information with imperfect channel state information,'' \emph{Entropy}, vol.~23, no.~12, 2021.

\bibitem{costa2015fsmc}
M.~Costa, S.~Valentin, and A.~Ephremides, ``On the age of channel information for a finite-state markov model,'' in \emph{IEEE International Conference on Communications (ICC)}, 2015, pp. 4101--4106.

\bibitem{costa2015on}
M.~Costa, S.~Valentin, and A.~Ephremides, ``On the age of channel state information for non-reciprocal wireless links,'' in \emph{IEEE International Symposium on Information Theory (ISIT)}, 2015.

\bibitem{hasslinger2008gilbert}
G.~Hasslinger and O.~Hohlfeld, ``The gilbert-elliott model for packet loss in real time services on the internet,'' in \emph{Measurement, Modelling and Evalutation of Computer and Communication Systems}, 2008, pp. 1--15.

\bibitem{bildea2015link}
A.~Bildea, O.~Alphand, F.~Rousseau, and A.~Duda, ``Link quality estimation with the gilbert-elliot model for wireless sensor networks,'' in \emph{IEEE International Symposium on Personal, Indoor, and Mobile Radio Communications (PIMRC)}, 2015, pp. 2049--2054.

\bibitem{wang1995finite}
H.~S. Wang and N.~Moayeri, ``Finite-state markov channel-a useful model for radio communication channels,'' \emph{IEEE Transactions on Vehicular Technology}, vol.~44, no.~1, pp. 163--171, 1995.

\bibitem{ross2014introduction}
S.~M. Ross, \emph{Introduction to Probability Models (11th Edition)}.\hskip 1em plus 0.5em minus 0.4em\relax Academic Press, 2014.

\bibitem{puterman1994markov}
M.~L. Puterman, \emph{Markov decision processes: discrete stochastic dynamic programming}.\hskip 1em plus 0.5em minus 0.4em\relax John Wiley \& Sons, 1994.

\bibitem{luo2025TCOM}
J.~Luo and N.~Pappas, ``Semantic-aware remote estimation of multiple {Markov} sources under constraints,'' \emph{IEEE Trans. Commun.}, May 2025, early access.

\end{thebibliography}
